\documentclass[12pt]{article}
\usepackage{amsmath,amssymb,graphicx}

\usepackage[utf8]{inputenc}

\usepackage[T1]{fontenc}

\DeclareRobustCommand\openone{\leavevmode\hbox{\small1\normalsize\kern-.33em1}}
\usepackage[numbers,comma,sort&compress]{natbib}

\newlength{\fighskip} \fighskip=2pt
\newlength{\figvskip} \figvskip=3pt

\let\Im\relax\DeclareMathOperator{\Im}{Im}

\newcommand{\typ}{\mathrm{typ}}
\newcommand{\hc}{\text{\textit{h.c.}}}

\newcommand{\ii}{\mathrm{i}}
\DeclareMathOperator{\Tr}{Tr}
\DeclareMathOperator{\Texp}{T\,exp}

\title{Dissipationless dynamics of randomly coupled spins at high temperatures}

\author{Lara Faoro$\,{}^{1,2}$, Lev Ioffe$\,{}^{2}$, Alexei Kitaev$\,{}^{3}$\\
\parbox[t]{14cm}{\centering\normalsize\it
$^{1}$ Laboratoire de Physique Théorique et Hautes Énergies, CNRS UMR 7589,
Universités Paris 6 et 7, 4 place Jussieu, 75252 Paris, Cedex 05,
France\\
$^{2}$ Department of Physics and Astronomy, Rutgers University, 136 Frelinghuysen Rd, Piscataway 08854, New Jersey, USA\\
$^{3}$ California Institute of Technology, Pasadena, CA, 91125, USA}
}

\date{\today}

\begin{document}
\maketitle

\begin{abstract}
We develop a technique to compute the high-frequency asymptotics of spin correlators in weakly interacting disordered spin systems. We show that the dynamical spin correlator decreases exponentially at high frequencies, $\left\langle SS\right\rangle_{\omega}\sim\exp(-\tau_{*}\omega)$ and compute the characteristic time $\tau^{*}$ of this dependence.  In a typical random configuration, some fraction of spins form strongly coupled pairs, which behave as two-level systems. Their switching dynamics is driven by the high-frequency noise from the surrounding spins, resulting in low-frequency $1/f$ noise in the magnetic susceptibility and other physical quantities. We discuss application of these results to the problem of susceptibility and flux noise in superconducting circuits at mK temperatures.
\end{abstract}

\section*{Introduction}

In many physical systems, the relevant degrees of freedom are discrete degenerate quantum variables such as spins (electron or nuclear) that interact weakly with each other. At sufficiently high temperatures, the spins are completely disordered and characterized by the trivial thermodynamic correlators. However, their dynamical properties remain interesting and important for various applications. The spins generally fluctuate at frequencies of the order of the typical interaction $J_{\typ}$.  For some purposes, though, one needs to know the asymptotics of the spin correlation function at much higher frequencies. In this regime, the fluctuations and dynamical response are due to the simultaneous excitation of many degrees of freedom and inherent nonlinearity of the system. The goal of this paper is to develop a formalism for the computation of the high-frequency spin correlator and to apply the result to the problem of magnetic (flux and inductance) noise at mK temperatures produced by paramagnetic spins located at the surface of superconductor.

Specifically, we consider a set of spins described by the Heisenberg Hamiltonian with random couplings at high temperatures, $T\gg J_{\typ}$.  We show that the spin correlator decays exponentially with frequency at $\omega\gg J_{\typ}$, i.e.\ $\left\langle SS\right\rangle_{\omega}\sim\exp(-c\omega/J_{\typ})$, and compute the coefficient $c$ in this formula.

The smallness of the high frequency fluctuations has an important consequence when the interaction, although of the order of $J_{\typ}$ for the typical pairs of spins, is much larger for a small but significant number of pairs. This happens, for instance, if the spins are positioned randomly while the interaction between them falls off fast with distance. Each strongly coupled pair has two states, the singlet and the triplet. Transitions between those states require the exchange of a large energy quantum, $\Delta E=J\gg J_{\typ}$ between the pair and the surrounding spins. The transition rate is proportional to the dynamical spin correlator at frequency $\Delta E$, which is exponentially small. As a result, the switching events are rare and produce noise with, as we show below, $1/f$ spectrum at very \emph{low frequencies}. At intermediate frequencies, the $1/f$ noise is due to spin diffusion~\cite{Faoro2008}.

Our arguments rely on two important assumptions. First, the condition on the temperature is stricter than in the high frequency case: we need that $T\gg J_{\typ}\ln(J_{\typ}/\omega)$ if the noise is measured at frequency $\omega$. The right-hand side of this inequality represents the transition energy $\Delta E$ of the relevant two-level systems; if it is too large, such systems are frozen. We also assume that there is no efficient energy  exchange mechanism between the two-level systems and an external thermal bath, e.g.\ electrons or phonons.

One application of our theory is the property of paramagnetic spins on the surface of superconducting aluminum. These spins, presumably located in the thin Al$_{2}$O$_{3}$ film or just at the metal-insulator interface, are believed to be responsible for the low frequency flux noise that limits the sensitivity of dc SQUIDs~\cite{Koch1983,Wellstood1987} and causes decoherence in superconducting flux qubits~\cite{Yoshihara2006,Kakuyanagi2007,Bialczak2007}. It is very likely that the coupling between the spins is due to RKKY interaction~\cite{Faoro2008}, so their dynamics is controlled by the Heisenberg Hamiltonian. A number of experiments~\cite{Wellstood1987,Sendelbach2008,Lanting2009} shows that the $1/f$ noise at very low frequencies is temperature-independent fot $T\sim 50\,\text{--}\,500$~mK. We take these data as evidence that the inter-spin interaction $J_{\typ}$ is so small that the spins remain in the high temperature regime in the whole experimental range. Finally, in the superconductor all electronic excitations are gapped, and direct interaction between the spins and phonons at low temperature is negligible. All these facts combined together indicate that the spins in the oxide layer on the surface of aluminium or similar superconductors are described by the model studied in this paper.

A lower bound on $J_{\typ}$ can be inferred from the recent measurements of the flux qubit relaxation rate~\cite{Bylander2011}. If we assume that the relaxation is mainly due to the interaction with the surface spin, then ${\Gamma_{\text{relax}}\propto\Im\langle SS\rangle_{\omega=\Delta E}}$, where $\Delta E$ is the energy difference between the two qubit states. The experimental data indicate the persistence of magnetic relaxation at very high frequencies, at least up to $1$~GHz. We conclude that the typical interaction between the surface spins is $J_{\typ}\gtrsim 1\,\text{GHz}\sim 50\,\text{mK}$.  This is consistent with the observed spin freezing at $\sim 50\,\text{mK}$ in some samples~\cite{Sendelbach2008} as well as a theoretical estimate of the RKKY interaction, see Section~\ref{sect:low_frequency}.

The theory developed in this paper has many other applications, besides the problem of magnetic noise in superconducting circuits. It can be applied to many problems in which the relaxation of high energy modes is possible only by the simultaneous creation of a very large number of low energy excitations. Some examples are the relaxation of inverted nuclear spin polarization in high magnetic fields and a similar problem of magnetization relaxation of cold atoms following a rapid quench~\cite{Barmettler2009} or $1/f$ magnetic noise observed in a low temperature calorimeter with superconducting particle absorbers~\cite{Enss2005}.

The paper is organized as follows: we first study high frequency spin fluctuations at infinite temperature; then we show how the high frequency noise generates low frequency noise in a model with a broad distribution of coupling parameters. We finally discuss the application of this model to the problem of flux noise in superconducting qubits.

\section{High frequency noise}
We consider a set of spins of magnitude $S$ described by the Heisenberg Hamiltonian,
\begin{equation}
H=-\sum_{j<k}J_{jk}\vec{S}_{j}\vec{S}_{k}
\label{Heis}
\end{equation}
with random couplings $J_{jk}$. We are primarily interested in the $S=1/2$ case because it is most relevant for the physical applications. Let us fix $j$ and consider the normalized correlation function for the spin $\vec{S}_{j}$ (with components $S_{j}^{a}$,\, $a=1,2,3$) in the time and frequency domains:
\begin{equation}
F(t) = \frac{1}{S(S+1)} \left\langle S_{j}^{c}(t)
S_{j}^{c}(0)\right\rangle,\qquad
\tilde{F}(\omega)=\int_{-\infty}^{+\infty}F(t)e^{\ii\omega t}\,dt.
\label{eq:F}
\end{equation}
Note that $\tilde{F}(\omega)$ is real and positive. This follows from its physical interpretation as the quantum transition probability under a small perturbation $V(t)=-h^{a}e^{-\ii\omega t}S_{j}^{a}+\hc$ (More abstractly, $\int_{-\infty}^{+\infty} \langle X^{\dag}(t)X\rangle\,dt\ge 0$ if the averaging is performed over any quantum state that commutes with the Hamiltonian.)

In general, the asymptotics of the Fourier transform at large $\omega$ are related to analytic properties of the original function. However, since $\tilde{F}(\omega)\ge 0$, it is sufficient to consider $F(t)$ on the imaginary axis: $F(\ii\tau)=\int_{-\infty}^{\infty}\tilde{F}(\omega) e^{\tau\omega}\frac{d\omega}{2\pi}$. If the integral converges for $0\le\tau<\tau_{*}$, then $\tilde{F}(\omega)$ should decay sufficiently fast, and the real-time correlator has an analytic continuation to the strip $0\le\Im t<\tau_{*}$. If, in addition, the integral diverges for $\tau>\tau_{*}$, then $\tilde{F}(\omega)\sim e^{-\tau_{*}\omega}$ for $\omega\to+\infty$. (Strictly speaking, this expression is valid if we smooth out some unimportant features, such as peaks at discrete frequencies.)

At infinite temperature, the operator average in~(\ref{eq:F}) is taken over the maximally mixed state, i.e.\ $\langle X\rangle=\Tr X /\Tr\openone$. The cyclic property of the trace implies that both variants of the correlation function are even and real. Therefore, the Taylor expansion of the real-time spin correlator contains only even terms:
\begin{gather}
F(t) = \frac{1}{S(S+1)}
\left\langle e^{iHt}S_{j}^{c}(0)e^{-iHt}S_{j}^{c}(0)\right\rangle
=\sum_{n=0}^{\infty}\frac{(-1)^{n}D_{2n}}{(2n)!}t^{2n},
\label{eq:Taylor}\\[3pt]
D_{2n} = \frac{(-1)^n}{S(S+1)}\lim_{t\rightarrow 0}
\left\langle\frac{d^{2n}}{dt^{2n}} S_{j}^{c}(t)
S_{j}^{c}(0)\right\rangle
=\int_{-\infty}^{+\infty}\tilde{F}(\omega)\omega^{2n}\frac{d\omega}{2\pi}.
\label{eq:D_2n}
\end{gather}
We are interested in the asymptotic behavior of the moments $D_{2n}$ for $n\rightarrow\infty$. Let us assume that the power series~(\ref{eq:Taylor}) has a finite convergence radius $\tau_{*}$. Then $F(t)$ has singularities at $t=\pm\ii\tau_{*}$, and the frequency representation of the correlation function decays as $e^{-\tau_{*}|\omega|}$ for large $\omega$. Our main goal is to show that $\tau_{*}$ is nonzero and finite, and to compute its value.

Physically, the spin correlation function does not completely vanish at high frequencies only due to nonlinearities that allow for combining the precession frequencies of a large number of spins. To understand it qualitatively, consider an infinite system in which each spin is coupled to $Z$ neighbors with interaction strength $J_{\typ}$. If an energy quantum $\omega\gg J_{\typ}$ is exchanged between a single spin and some external object, it has to be distributed between other spins and divided into portions of the order of $J_{\typ}$. The most efficient processes of this kind are cascades that correspond to tree-like diagrams and do not involve any back reaction. Summing over such diagrams is equivalent to the mean-field approximation. It is generally applicable for $Z\gg1$, but will also produce qualitatively correct results if the energy has many paths to escape, for example, on a tree with $Z>2$.

Thus, we assume that each spin $j$ precesses in the local exchange field $\vec{h}_{j}(t)=\sum_{k}J_{jk}\vec{S}_{k}(t)$ produced by the spins surrounding it. The fluctuating field $\vec{h}(t)=\vec{h}_{j}(t)$ is characterized by the correlation function
\begin{equation}
\langle h^{a}(t)h^{b}(0)\rangle=g^{2}F(t)\delta^{ab},
\label{hh}
\end{equation}
where $a$, $b$ are spatial indices ($a,b=1,2,3$) and $g^{2}=\frac{S(S+1)}{3}\overline{\sum_{k}J_{jk}^{2}}$. Using the mean-field approach, we assume that $\vec{h}(t)$ is Gaussian. The relevant energy scale in this model (corresponding to $J_{\typ}$ in the qualitative discussion above) is given by $g$. In this approximation, the time evolution of each spin simplifies to $d\vec{S}/dt=\vec{S}\times\vec{h}(t)$, or, in a matrix form,
\begin{equation}
dS^{c}/dt=[\hat{h}(t)]_{ca}S^{a}.
\label{eq:ds/dt}
\end{equation}
Here $\hat{h}$ is the matrix corresponding to the cross-product with $\vec{h}$, and $[\hat{h}]_{ca}$ are its elements:
\begin{equation}
\hat{h}=\left(\begin{array}{ccc}
    0 & h^{3} & -h^{2}\\
    -h^{3} & 0 & h^{1}\\
    h^{2} & -h^{1} & 0\end{array}\right),\qquad
[\hat{h}]_{ca}=\epsilon_{cab}h^{b}.
\label{eq:matr}
\end{equation}
We solve equation~(\ref{eq:ds/dt}) in a symbolic form, $S^{c}(t)=\left[\Texp\int_{0}^{t}\hat{h}(t)\,dt\right]_{ca}S^{a}(0)$ and average over the thermal state of the spin: $\langle S^{a}S^{c}\rangle=\frac{S(S+1)}{3}\delta_{ac}$. Thus, the spin correlator is expressed as an average over the fluctuating fields:
\begin{equation}
F(t) =\frac{1}{3}\left\langle\Tr\left(\Texp\int_{0}^{t}\hat{h}(t)\,dt
\right)\right\rangle.
\label{adjrep}
\end{equation}
Using equations~(\ref{adjrep}) and~(\ref{hh}), we recursively compute the first few moments $D_{2n}$:
\begin{equation}
\begin{aligned}
D_{0} & =1,\\
D_{2} & = 2g^{2}D_{0},\\
D_{4} & = 2g^{2}D_{2}+10g^{4}D_{0}^{2},\\
D_{6} & = 2g^{2}D_{4}+48g^{4}D_{2}D_{0}+70g^{6}D_{0}^{3}.
\end{aligned}
\label{rr}
\end{equation}
It is unfortunately difficult to pursue this computation much further and find the asymptotic behavior of the moments. Nevertheless, these recursive formulas give an exact expansion of the spin correlation function in lower orders:
\begin{equation} F(t)=1-(gt)^{2}+\frac{7}{12}(gt)^{4}-\frac{97}{360}(gt)^{6}+O(gt)^{8}.
\label{firstord}
\end{equation}

Let us describe a general method for the calculation of the moments. It can be used to derive equations~(\ref{rr}) and continue to higher orders, though we could not obtain a closed-form expression for $D_{2n}$. We first use the equation of motion~(\ref{eq:ds/dt}) to express the $2n$-th derivative in Eq.~(\ref{eq:D_2n}):
\begin{equation}
\frac{d^{2n}S^{c}}{dt^{2n}}
=\sum_{k=1}^{{2n}}\binom{2n-1}{k-1}\,[\hat{h}^{(k-1)}]_{ca}
\frac{d^{2n-k}S^{a}}{dt^{2n-k}},
\label{2nthderiv}
\end{equation}
where $\hat{h}^{(k)}=d^{k}\hat{h}/dt^{k}$. By applying Eq.~(\ref{2nthderiv}) repeatedly, we obtain an expansion
\begin{equation}
\frac{d^{2n}S^{c}}{dt^{2n}} = \sum_{m\ge2}\,
\sum_{\substack{k_1,\dotsc,k_m\ge1\\k_1+\dotsb+k_m=2n}}
\hspace{-4pt}C_{k_{1},\dotsc,k_{m}}\, [\hat{h}^{(k_{1}-1)}]_{ca_{1}}\dotsm
[\hat{h}^{(k_{m}-1)}]_{a_{m-1},a_{m}}\, S^{a_{m}} S^{c}
\label{eq:derivCk}
\end{equation}
with the coefficients
\begin{equation}
C_{k_{1},\dotsc,k_{m}} =\frac{(2n-1)!}
{(2n-k_{1})\dotsm(2n-k_{1}-\dotsb-k_{m-1})\:(k_{1}-1)!\dotsm(k_{m}-1)!}.
\label{eq:Ck}
\end{equation}
Recall that the spin operators in~(\ref{eq:derivCk}) pertain to site $j$, and the fields $\hat{h}$ are linear combinations of adjacent spins. The field derivatives $\hat{h}^{(k)}$ can be expressed in terms of further neighbors. So far all the calculations have been exact. Now we employ the mean-field approximation and average over the thermal state of the spin, and then over the Gaussian fields. Thus,
\begin{equation}
D_{2n} = \,\frac{(-1)^{n}}{3}\,\sum_{m\ge2}
\sum_{\substack{k_1,\dotsc,k_m\ge1\\k_1+\dotsb+k_m=2n}}
\hspace{-4pt} C_{k_{1},\dots,k_{m}}
\left\langle\Tr\bigl(\hat{h}^{(k_{1}-1)} \hat{h}^{(k_{2}-1)}
\dotsm\hat{h}^{(k_{m}-1)}\bigr)\right\rangle.
\label{ex}
\end{equation}
The Gaussian average in this formula can be evaluated using Wick's theorem and the explicit form of the second-order correlator:
\begin{equation}
\Bigl\langle [h^{(k)}]_{ab}\,[h^{(l)}]_{cd} \Bigr\rangle 
=g^{2}\, (-1)^{\frac{k-l}{2}}D_{k+l}\,
\bigl(\delta_{ac}\delta_{bd}-\delta_{ad}\delta_{bc}\bigr).
\label{gaussav}
\end{equation}

\section{Estimates for $\tau_{*}$}
The difficulty in carrying out the above calculation lies in the matrix nature of the fields $\hat{h}$. We now introduce two different approximations, which are expected to shift the answer in opposite directions by overestimating or underestimating the moments $D_{2n}$, thus giving a lower and an upper bound for $\tau_{*}$. Remarkably, we find that the two bounds are quite close in value.
%As we will see, the difference is surprisingly small.

In our first approximation, we ignore the noncommutativity of the matrices $\hat{h}^{(k)}$, or equivalently, $\hat{h}(t)$ for different values of $t$. More exactly, we symmetrize over all permutations of those matrices. Let us briefly explain how to use this recipe for equation~(\ref{ex}), although it is easier to start almost from scratch as we will do later. By symmetrizing the combinatorial coefficients~(\ref{eq:Ck}), we obtain the expression $\frac{1}{m!}\frac{(2n)!}{k_1!\dotsm k_m!}$. Then we consider an individual Wick pairing for the product $\hat{h}^{(k_{1}-1)}\dotsm\hat{h}^{(k_{m}-1)}$ and relate it to the corresponding paring for the $m$-th power of $\hat{h}$. The ratio is given by the scalar coefficient $(-1)^{\frac{k-l}{2}}D_{k+l}$ in Eq.~(\ref{gaussav}); there is one such factor for each pair. Summing over all Wick pairings, we get $\bigl\langle\Tr\hat{h}^m\bigr\rangle$ times a certain number. It is easy to show that $\bigl\langle\Tr\hat{h}^{2r}\bigr\rangle =\bigl(-g^2\bigr)^{r}\frac{(2r+1)!}{2^{r-1}r!}$ for $r>0$. Eventually, Eq.~(\ref{ex}) becomes:
\begin{equation}
\frac{D_{2n}}{(2n)!} =\,\frac{2}{3}\,\sum_{r\ge1}\frac{(2r+1)}{r!}g^{2r}
\hspace{-6pt}\sum_{\substack{q_1,\dotsc,q_r\ge2\\q_1+\dotsb+q_r=2n}}
\hspace{-4pt}\frac{D_{q_{1}-2}}{q_{1}!}\dotsm \frac{D_{q_{r}-2}}{q_{r}!}\,.
\label{rec}
\end{equation}
The solution to this recurrence yields the following expansion of $F(t)$ in lower orders:
\begin{equation} F(t)=1-(gt)^{2}+\frac{7}{12}(gt)^{4}-\frac{99}{360}(gt)^{6}+...\;
\label{firstordL}
\end{equation}
Notice the subtle difference from the exact solution~(\ref{firstord}). We believe that also in higher orders, the noncommutativity of the field matrices results in smaller values of $D_{2n}$ than the ones obtained by the present method.

The same approximation can be applied directly to Eq.~(\ref{adjrep}) if we replace the time ordered exponential with an ordinary exponential:
\begin{equation}
F(t) =\frac{1}{3}
\left\langle\Tr\bigl(\exp \hat{H}(t)\bigr)\right\rangle
=\frac{1}{3}\left\langle 1+ 2\cos\bigl|\vec{H}(t)\bigr|\right\rangle,\qquad
\vec{H}(t)=\int_{0}^{t}\vec{h}(t)\,dt,
\label{eq:F(t)cos}
\end{equation}
where $\hat{H}$ is the matrix associated with vector $\vec{H}$. The vector components $H^{a}(t)$\, ($a=1,2,3$) are Gaussian fields with the following second moment:
\begin{gather}
\langle H^{a}(t)H^{b}(t)\rangle=2g^{2}G(t)\delta^{ab},\\
G(t)=\int_{0}^{t}(t-t_1)F(t_1)\,dt_1.
\label{eq:G}
\end{gather}
Higher moments $\langle{|\vec{H}|^{2n}}\rangle$ (for a given $t$) are calculated directly by integrating over $\vec{H}$. Using the notation $x=|\vec{H}|/(2g^{2}G(t))^{1/2}$, we find that
\begin{equation}
\langle x^{2n}\rangle
=\frac{\int_{0}^{\infty}x^{2n}\,e^{-x^2/2}\,x^{2}dx}
{\int_{0}^{\infty}e^{-x^2/2}\,x^{2}dx}
=2^{n}\frac{\Gamma(n+3/2)}{\Gamma(3/2)}
=\frac{(2n+1)!}{2^{n}n!}.
\end{equation}
Now we calculate the average in Eq.~(\ref{eq:F(t)cos}) by expanding 
$\cos|\vec{H}(t)|$ in powers of $|\vec{H}|^2$. The result is:
\begin{equation}
F(t)=\frac{1}{3}\Bigl(1+ 2\bigl(1-2g^{2}G(t)\bigr)\,e^{-g^{2}G(t)}\Bigr).
\label{finL}
\end{equation}
This equation needs to be solved in conjunction with Eq.~(\ref{eq:G}).

To compute $F(\ii\tau)$, we formally replace $t$ with $\tau$ and $-g^2$ with $+g^2$. The numerical solution exhibits singularities at $\tau=\pm\tau_{*}$ for $g\tau_{*}\approx1.74$. Since the moments $D_{2n}$ have been overestimated, this gives a lower bound for the true value of $\tau_{*}$.

We now make a different approximation to matrix products, exaggerating the noncommutativity. Specifically, we replace the physical $S=1/2$ spin for the $SU(2)$ group with an $SU(N)$ spin and consider the limit $N\to\infty$. Let us first introduce the generators of $U(N)$: these are $N\times N$ matrices $\sigma^{\alpha\beta}$\, ($\alpha,\beta=1,\dotsc,N$) with the elements  $[\sigma^{\alpha\beta}]_{\mu\nu}=\delta_{\alpha\mu}\delta_{\beta\nu}$. The $SU(N)$ spin is defined by the traceless matrices
\begin{equation}
\tilde{\sigma}^{\alpha\beta}
=\sigma^{\alpha\beta}-\frac{1}{N}\delta^{\alpha\beta}\openone.
\label{generators}
\end{equation}
The generalized Heisenberg interaction is given by the operator $\frac{1}{N}\tilde{\sigma}_{j}^{\alpha\beta}\tilde{\sigma}_{k}^{\beta\alpha}$ with the matrix elements $\frac{1}{N}\bigl[\tilde{\sigma}^{\alpha\beta}\bigr]_{\gamma\delta} \bigl[\tilde{\sigma}^{\beta\alpha}\bigr]_{\mu\nu} =\frac{1}{N}\Bigl(\delta_{\gamma\nu}\delta_{\delta\mu} -\frac{1}{N}\delta_{\gamma\delta}\delta_{\mu\nu}\Bigr)$. Note that if we let $\tilde{\sigma}^{\alpha\beta}$ and $\tilde{\sigma}^{\beta\alpha}$ act on the same spin (by contracting over a pair of indices, $\delta=\mu$), we get $1-\frac{1}{N^{2}}$ times the identity matrix. For $N=2$, this matches the value of $S^{a}S^{a}=S(S+1)$. 

Thus, the full Hamiltonian has the form
\begin{equation}
H=-\frac{1}{N}\sum_{j<k}J_{jk}
\tilde{\sigma}_{j}^{\alpha\beta}\tilde{\sigma}_{k}^{\beta\alpha}.
\label{HeisN}
\end{equation}
Each spin $j$ experiences the effective field $\tilde{h}_{j}^{\alpha\beta}(t) =\frac{1}{N}\sum_{k} J_{jk} \tilde{\sigma}_{k}^{\alpha\beta}$. In the mean-field approximation, we have:
\begin{equation}
\bigl\langle\tilde{h}^{\alpha\beta}(t)
\tilde{h}^{\beta'\alpha'}\!(0)\bigr\rangle
   =g^{2}\bigl\langle\tilde{\sigma}^{\alpha\beta}(t)
\tilde{\sigma}^{\beta'\alpha'}\!(0)\bigr\rangle
=\frac{g^{2}}{N} F_{N}(t)
\left(\delta^{\alpha\alpha'}\delta^{\beta\beta'}
-\frac{1}{N}\delta^{\alpha\beta}\delta^{\beta'\alpha'}\right),
\label{gaussN}
\end{equation}
where $g^2=\frac{1}{N^{2}}\overline{\sum_{k}J_{jk}^{2}}$, and the spin correlator is normalized so that $F_N(0)=1$. Note that the last term in~(\ref{gaussN}) has no effect when we consider the action of the field on the spin, because $\delta^{\alpha\beta}\tilde{\sigma}^{\beta\alpha}=0$.

\begin{figure}[t]
\center
\includegraphics[width=4in]{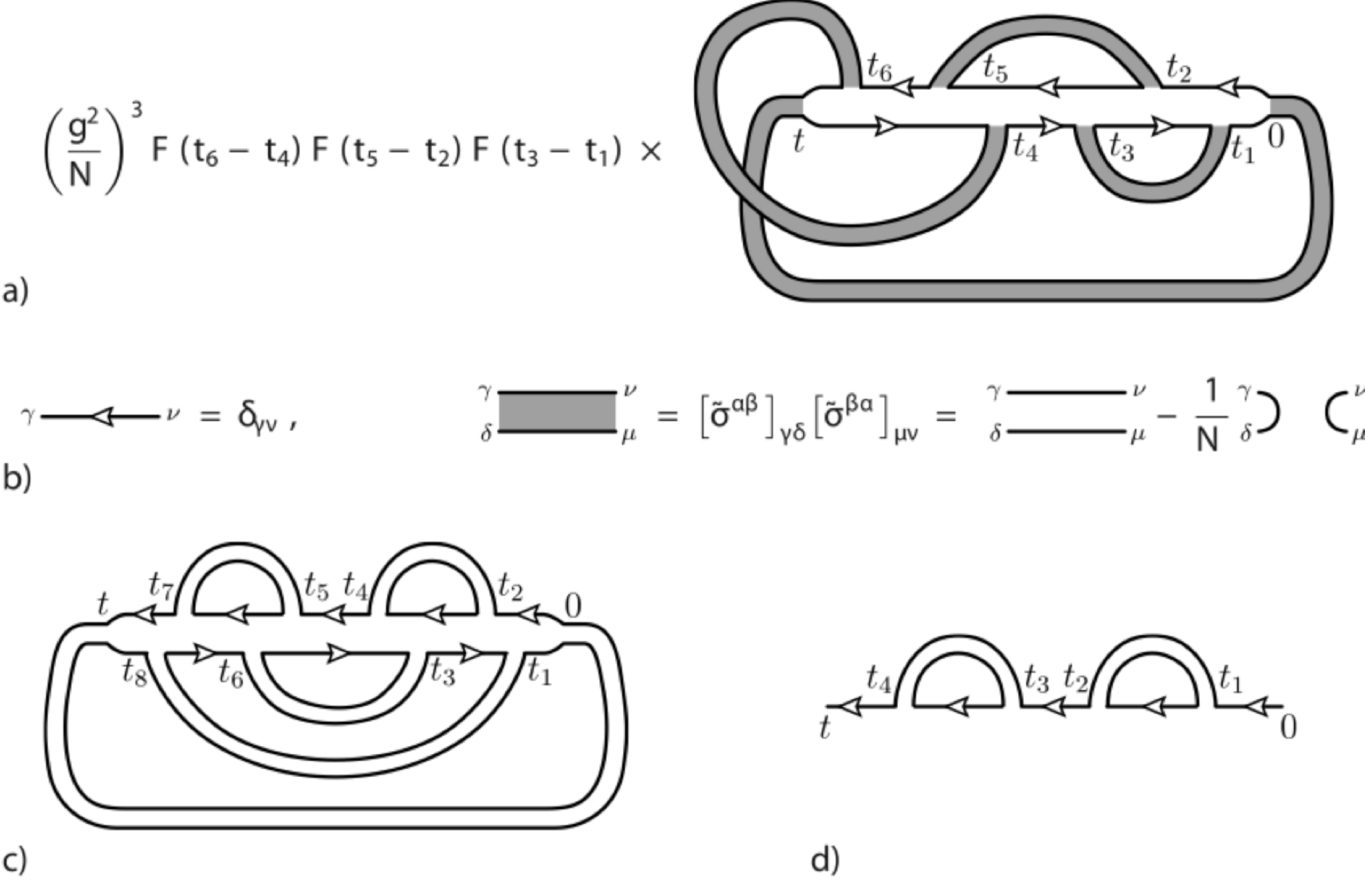}
\caption{(a)~A generic term in the diagrammatic expansion of 
$\bigl\langle\Tr\bigl( U_{N}^{\dag}(t)\,\tilde{\sigma}^{\alpha\beta}\, U_{N}(t)\,\tilde{\sigma}^{\beta\alpha}\bigr)\bigr\rangle$ (for an arbitrary $N$) using equations~(\ref{gaussN}) and~(\ref{evolN}). The arrows indicate $U_{N}(t)$ (the propagation from $0$ to $t$) and $U_{N}^{\dag}(t)$ (moving backward in time); the shaded ribbons correspond to field correlators.\; (b)~The exact definition of lines and ribbons. Note that that they do not depend on time: the factors $F(t_k-t_j)$ are coefficients in front of the diagram.\; (c)~A planar diagram: one of the leading terms in the $N\to\infty$ limit.\; (d)~A diagram for the function $f(t)$ defined in~(\ref{eq:f}).}
\label{fig:diaggen}
\end{figure}

The calculation of
\begin{equation}
F_{N}(t)=\frac{1}{N^{2}-1}\,\left\langle\Tr\bigl( U_{N}^{\dag}(t)\,\tilde{\sigma}^{\alpha\beta}\, U_{N}(t)\,\tilde{\sigma}^{\beta\alpha}\bigr)\right\rangle,\qquad
U_{N}(t) =\Texp\int_{0}^{t}\ii\,\tilde{h}^{\alpha\beta}(t_{1})\,
\tilde{\sigma}^{\beta\alpha}\, dt_{1}
\label{evolN}
\end{equation}
can be performed using the standard diagrammatic technique for matrix models, see Fig.~\ref{fig:diaggen}\,a,b. The ribbons in the diagrams may actually be changed to double lines. To justify this step, let us express $\tilde{\sigma}^{\alpha\beta}$ in terms of $\sigma^{\alpha\beta}$ using~(\ref{generators}). The result is:  $\bigl\langle\Tr\bigl( U_{N}^{\dag}(t)\,\tilde{\sigma}^{\alpha\beta}\, U_{N}(t)\,\tilde{\sigma}^{\beta\alpha}\bigr)\bigr\rangle =\bigl\langle\Tr\bigl( U_{N}^{\dag}(t)\,\sigma^{\alpha\beta}\, U_{N}(t)\,\sigma^{\beta\alpha}\bigr)\bigr\rangle -1$, which shows that the ribbon between $0$ and $t$ can be changed as stated (up to the $-1$ term). We can likewise replace the $\tilde{\sigma}^{\beta\alpha}$ that enter the expressions for $U_{N}(t)$, $U_{N}^{\dag}(t)$. The contributions of the unit operator in~(\ref{generators}) to $U_{N}(t)$ and to $U_{N}^{\dag}(t)$ cancel each other, therefore the remaining ribbons are also equivalent to double lines.

A great further simplification occurs in the limit of large $N$~\cite{tHooft1974}. The leading contribution to $F_{N}(t)$ comes from diagrams that maximize the number of loops for a given expansion order in $g^2/N$. Such diagrams are planar, and the time evolution factors $U_{N}(t)$, $U_{N}^{\dag}(t)$ separate; an example is shown in Fig.~\ref{fig:diaggen}\,c. Thus, in the limit $N\to\infty$ equations~(\ref{gaussN}) and~(\ref{evolN}) become:
\begin{gather}
\langle h^{\alpha\beta}(t)h^{\beta^{\prime}\alpha^{\prime}}(0)\rangle
  =\frac{g^{2}}{N}F_{\infty}(t)\,
\delta^{\alpha\alpha^{\prime}}\delta^{\beta\beta^{\prime}},\\[3pt]
F_{\infty}(t)=f^{2}(t),\qquad
f(t)\openone
  =\left\langle \Texp\int_{0}^{t}\ii\,
h^{\alpha\beta}(\tau)\sigma^{\beta\alpha}\, d\tau\right\rangle.
\label{eq:f}
\end{gather}
The ``spin propagator'' $f(t)$ can be computed by summing up planar diagrams. Specifically, $f(t)$ is represented by a single line going from $0$ to $t$ and dressed with double lines (the field correlators) attached on one side; these lines do not intersect, see Fig.~\ref{fig:diaggen}\,d. This diagrammatic expression leads to the Dyson equation:
\begin{equation}
f(t)=1-g^{2}\int_{t>t_{2}>t_{1}>0} dt_{2} dt_{1}
f^{3}(t_{2}-t_{1})\, f(t_{1}).
\label{sid}
\end{equation}

Curiously, the Taylor expansion for $F_{\infty}(t)=f^{2}(t)$ coincides with the result~(\ref{firstord}) for $F_{2}(t)$ up to the sixth order. The numerical solution of Eq.~(\ref{sid}) exhibits singularities at $t=\pm\ii \tau_{*}$ for $g\tau_{*}\approx1.78$. Thus, the exact value of $\tau_{*}$ for the original problem should be between $1.74/g$ and $1.78/g$. Recall that this number is the exponent in the expression for the high frequency noise spectrum: $\langle h_{\omega}^{2}\rangle\propto e^{-\tau_{*}|\omega|}$.

\section{Low frequency noise}\label{sect:low_frequency}
The high frequency fluctuations of local effective fields drive the switching dynamics of strongly coupled spin pairs, and thus determine the noise at low frequencies (provided the number of such pairs is significant). We will find the spectrum of the low frequency noise in a specific model, which has been previously discussed in~\cite{Faoro2008}.

Let us consider randomly distributed spins with the two-dimensional density $\rho_{2D}$ on the surface of a superconductor. The coupling of each spin to the conduction electrons is described by the Kondo term in the Hamiltonian: ${\mathcal{J}_j\,S_j^{\alpha}s^{\alpha}(\vec{r}_j)}$, where ${s^\alpha(\vec{r})=\frac{1}{2}\left[\sigma^{\alpha}\right]_{\mu\nu} \hat{\psi}_{\mu}^{\dag}(\vec{r})\hat{\psi}_{\nu}(\vec{r})}$. The relative coupling strength can be characterized by the dimensionless parameter
\begin{equation}
\lambda=\mathcal{J}\nu=\bigl(\ln(E_{F}/T_{K})\bigr)^{-1},
\end{equation}
where $\nu$ is the electron density of states per spin projection and $T_K$ is the Kondo temperature. The RKKY interaction between two spins can be expressed as follows:
\begin{equation}
J_{jk}=-\frac{\mathcal{J}_j\mathcal{J}_k\nu}{8\pi r_{jk}^{3}}\cos(2k_{F}r_{jk})
=J(r_{jk})\cos\phi_{jk},\qquad
J(r)=\frac{\lambda^2}{8\pi\nu}\,r^{-3},
\label{eq:RKKY}
\end{equation}
where $\phi_{jk}$ is random, and $r_{jk}$ is assumed to be much smaller than the coherence length. The surface is usually covered by an insulator and thus can carry spins with arbitrary coupling to the conduction electrons, ranging from $\lambda\sim1$ down to $0$. However, only the spins with $T_{K}\lesssim T_{c}$ are active, the other being quenched by the Kondo effect. This imposes the constraint $\lambda\lesssim0.1$. The typical distance between the spins is $r_{\typ}=\rho_{2D}^{-1/2}$, which defines the main parameter of our model: $g\sim J(r_{\typ})=\frac{\lambda^{2}}{4\pi\nu}\,\rho_{2D}^{3/2}$.  Spins coupled with this strength form an infinite cluster, whereas more strongly coupled pairs are rare. If we use the values $\nu\approx17\,\text{eV}^{-1}\text{nm}^{-3}$ (the DOS in aluminum) and $\rho_{2D}\sim0.4\,\text{nm}^{-2}$ (the typical spin density measured in several types of films~\cite{Sendelbach2008,Moler2009}), we get the estimate $g\lesssim 50\,\text{mK}$. The upper bound corresponds to the spins with the largest value of $\lambda$. There are some indications, both experimental~\cite{Moler2009} and theoretical~\cite{Clarke2009}, suggesting that most localized spins arise from surface bound state. Therefore we can expect the actual value of $g$ to be close to the upper bound.

The density of pairs with interaction $J\gg g$ can be found by calculating the probability that a given spin has a neighbor at the corresponding distance. Such pairs constantly switch between the singlet and the triplet states at a characteristic rate $\Gamma\sim g\, e^{-\tau_{*}J}=g\, e^{-cJ/g}$, where $c$ is a numerical constant. We estimate the density of pairs in terms of $\Gamma$:
\begin{equation}
\frac{d\rho}{\rho_{2D}}=\frac{1}{2}\,\rho_{2D}\,2\pi r\,dr
\sim (J/g)^{-5/3}\,d(J/g)
\approx \bigl(\ln(g/\Gamma)\bigr)^{-5/3}\frac{d\Gamma}{c\,\Gamma}.
\end{equation}
Each pair generates a random telegraph noise in any quantity $x$ that is proportional to the number of pairs in the triplet state; for example, $x$ can be the magnetic susceptibility. This noise has a Lorentzian spectrum with width $\Gamma$, but, for the purpose of a crude estimate, we may assume that it is concentrated at frequency $\omega=\Gamma$. Therefore, the total noise spectrum is proportional to the previously calculated distribution of $\Gamma$:
\begin{equation}
\left\langle (\delta x)_{\omega}^{2}\right\rangle 
\propto\frac{1}{\omega\ln^{5/3}(g/\omega)}.
\label{eq:noise spectrum}
\end{equation}

Thus, the magnetic susceptibility noise has a roughly $1/f$ spectrum at low frequencies. In the presence of a constant magnetic field, it translates to a $1/f$ flux noise. Some experimental data indicate a spontaneous breaking of the time reversal symmetry in spin systems at low temperatures \cite{Sendelbach2009}. Although the exact nature of the symmetry-breaking order parameter is unclear, it is most likely to exert local fields on the pairs of spins that undergo the slow fluctuations. In this case, the triplet state is split in energy and produce a net magnetization, observed as a flux noise.

\section*{Acknowledgments}
We are grateful to R. McDermott for the discussion of the experimental situation. This work was supported, in part, by Triangle de la physique 2007-36, ANR-06-BLAN-0218, ECS-0608842, ARO W911-09-1-0395, DARPA HR0011-09-1-0009.  A.K.\ acknowledges funding by the Institute of Quantum Information under NSF grant no.\ PHY-0803371.

\bibliographystyle{aipproc-arxiv}

\end{document}